\documentclass{article}

\usepackage[preprint]{neurips_2021}

\usepackage[utf8]{inputenc} % allow utf-8 input
\usepackage[T1]{fontenc}    % use 8-bit T1 fonts
\usepackage{hyperref}       % hyperlinks
\usepackage{url}            % simple URL typesetting
\usepackage{booktabs}       % professional-quality tables
\usepackage{amsfonts}       % blackboard math symbols
\usepackage{nicefrac}       % compact symbols for 1/2, etc.
\usepackage{microtype}      % microtypography
\usepackage{xcolor}         % colors

\usepackage{algorithm}
\usepackage{algpseudocode}

\newtheorem{theorem}{Theorem}[section]
\newtheorem{defi}[theorem]{Definition}

\usepackage{booktabs}
\usepackage{multirow}
\usepackage{siunitx}
\usepackage{amssymb}
\usepackage{booktabs}       % professional-quality tables
\usepackage{amsfonts}       % blackboard math symbols
\usepackage{nicefrac}       % compact symbols for 1/2, etc.
\usepackage{microtype}      % microtypography
\usepackage{amsmath}
\usepackage{graphicx}
\usepackage{multirow}
\usepackage{booktabs}
\usepackage{wrapfig}

\title{Highly Accurate Disease Diagnosis and \\ Highly Reproducible Biomarker Identification \\ with PathFormer}

\author{%
  Zehao Dong\textsuperscript{1}  Qihang Zhao\textsuperscript{8}, Philip R.O. Payne \textsuperscript{2} Michael A Province \textsuperscript{5} Carlos Cruchaga \textsuperscript{6} \\
  \textbf{Muhan Zhang \textsuperscript{4} Tianyu Zhao\textsuperscript{7}
  Yixin Chen\textsuperscript{1} Fuhai Li\textsuperscript{2,3, *}} \\
  \textsuperscript{1} Department of Computer Science $ \&$ Engineering, Washington University in St. Louis \\
   \textsuperscript{4} Institute for Artificial Intelligence, Peking University \\
  \textsuperscript{2} Institute for Informatics (I2), \textsuperscript{3} Department of Pediatrics, \textsuperscript{5} Department of Genetics, \\ \textsuperscript{6} Department of Psychiatry, \textsuperscript{7} Department of Radiation Oncology \\ Washington University School of Medicine, Washington University in St. Louis \\
  \textsuperscript{8} Department of Health Technology and Informatics, The Hong Kong Polytechnic University\\
  * Correspondence: Fuhai.Li@wustl.edu
}

\begin{document}

\maketitle

\begin{abstract}
   Biomarker identification is critical for precise disease diagnosis and understanding disease pathogenesis in omics data analysis, like using fold change and regression analysis. Graph neural networks (GNNs) have been the dominant deep learning model for analyzing graph-structured data. However, we found two major limitations of existing GNNs in omics data analysis, i.e., limited-prediction/diagnosis accuracy and limited-reproducible biomarker identification capacity across multiple datasets. The root of the challenges is the unique graph structure of biological signaling pathways, which consists of a large number of targets and intensive and complex signaling interactions among these targets. To resolve these two challenges, in this study, we presented a novel GNN model architecture, named PathFormer, which systematically integrate signaling network, priori knowledge and omics data to rank biomarkers and predict disease diagnosis. In the comparison results, PathFormer outperformed existing GNN models significantly in terms of highly accurate prediction capability (~30$\%$ accuracy improvement in disease diagnosis compared with existing GNN models) and high reproducibility of biomarker ranking across different datasets. The improvement was confirmed using two independent Alzheimer’s Disease (AD) and cancer transcriptomic datasets. The PathFormer model can be directly applied to other omics data analysis studies.

\end{abstract}
\section{Introduction}
Due to the advent of next generation sequencing (NGS) and high throughput technologies, large-scale and personalized omics data have been being generated. The analysis of the omics datasets has uncovered many novel disease-associated targets. However, for most of diseases, the complex and mysterious disease pathogenesis remains unclear yet. In the omics data analysis, biomarker or target identification is critical for precise disease diagnosis and understanding the disease pathogenesis in omics data analysis, like using fold change and regression analysis. For example, the fold change plus p-value via statistical analysis, or regression analysis are the widely used models to rank targets, followed by functional analysis of the top-ranked targets. However, these approaches cannot model the signaling interactions among these individual targets/proteins. On the other hand, signaling networks, like signaling pathways and protein-protein interactions (PPIs), which formulate the genome-wide association/interactions of multiple genes, are ubiquitous in various bioinformatical applications, including drug synergy prediction \cite{podolsky2011combination,hopkins2008network},  DVH prediction \cite{dong2024dosegnn}, Alzheimer’s disease (AD) detection \cite{song2019graph,qin2022aiding}, cancer classification \cite{lu2003cancer,viale2012current,amrane2018breast}, etc. The network-based analysis can identify the stable network module biomarkers or hub genes that can help understand the contributions of gene sets of pathways to disease phenotypes. 

Signaling networks are one type of essentially graph modality. Nowadays, deep learning models are in great demand to analyze signaling networks in a computational way. Various deep models \cite{yang2014gene, horvath2008geometric,song2015multiscale} have been proposed to predict disease phenotype from gene expressions, yet they did not consider the interactions between genes, thus failing to capture the joint role of multiple genes in determining the phenotypic variability. Graph neural networks (GNNs) \cite{gilmer2017neural, kipf2016semi, scarselli2008graph, Velickovic2018GraphAN, dong2022pace} are the dominant architecture for modeling graph-structured data and have achieved impressive performance on analysis tasks over various graphs, including social networks, molecules and circuits \cite{berg2017graph, bian2020rumor, dong2023cktgnn}. Though GNNs simultaneously encode the gene expression profiles and genetic interactions, several drawbacks limit their potential in the real-world bioinformatical applications. (1) First, current GNNs always exhibit subpar performance in predicting the disease phenotype. For instance, in the Alzheimer’s disease (AD) classification task where GNNs are required to distinguish AD samples from controls, we observe that the classification accuracies of existing/dominant GNNs are close to 0.6, which are slightly better than random guesses (limited diagnosis accuracy). (2) Second, current GNNs fail to provide interpretable results of biological meaning. A common interpretation pipeline is to extract gene subset from the input gene network, then gene set enrichment analysis (GSEA) can help researchers identify key biological pathways and processes. Though GNN architectures, such as SortPool and GAT, provide ways to rank the nodes’ contribution to select the gene subset, these techniques are not robust nor disease specific (limited-reproducible target ranking). 

The unique graph structure of biological signaling pathways, which consists of a large number of targets and intensive and complex signaling interactions among these targets, are believed as the root of the challenges. As seen in Figure ~\ref{fig:Fig3}-a,b, we mathematically characterized the limitations of existing GNNs in gene network representation learning. Specifically, compared to graphs in the popular benchmark graph datasets, gene networks usually contain thousands of genes/nodes, many of which has the extremely large node centrality. Such properties cause the subpar prediction performance of dominant GNNs from two aspects: (1) Dominant GNNs suffer the over-squashing problem \cite{xu2018powerful,alon2020bottleneck} for graphs with large average node degree/centrality; (2) Dominant expressive GNNs, such as subgraph-based GNNs and high-order GNNs \cite{morris2019weisfeiler, grohe2021logic}, have the space/time complexity issue when applied to large-scale graphs like gene networks. Thus, novel GNN models are needed for the signaling network-based omics data analysis for disease diagnosis and biomarker detection. 

\begin{figure}[t]
\begin{center}
%\vskip -0.15in
\centerline{\includegraphics[width=0.99\textwidth]{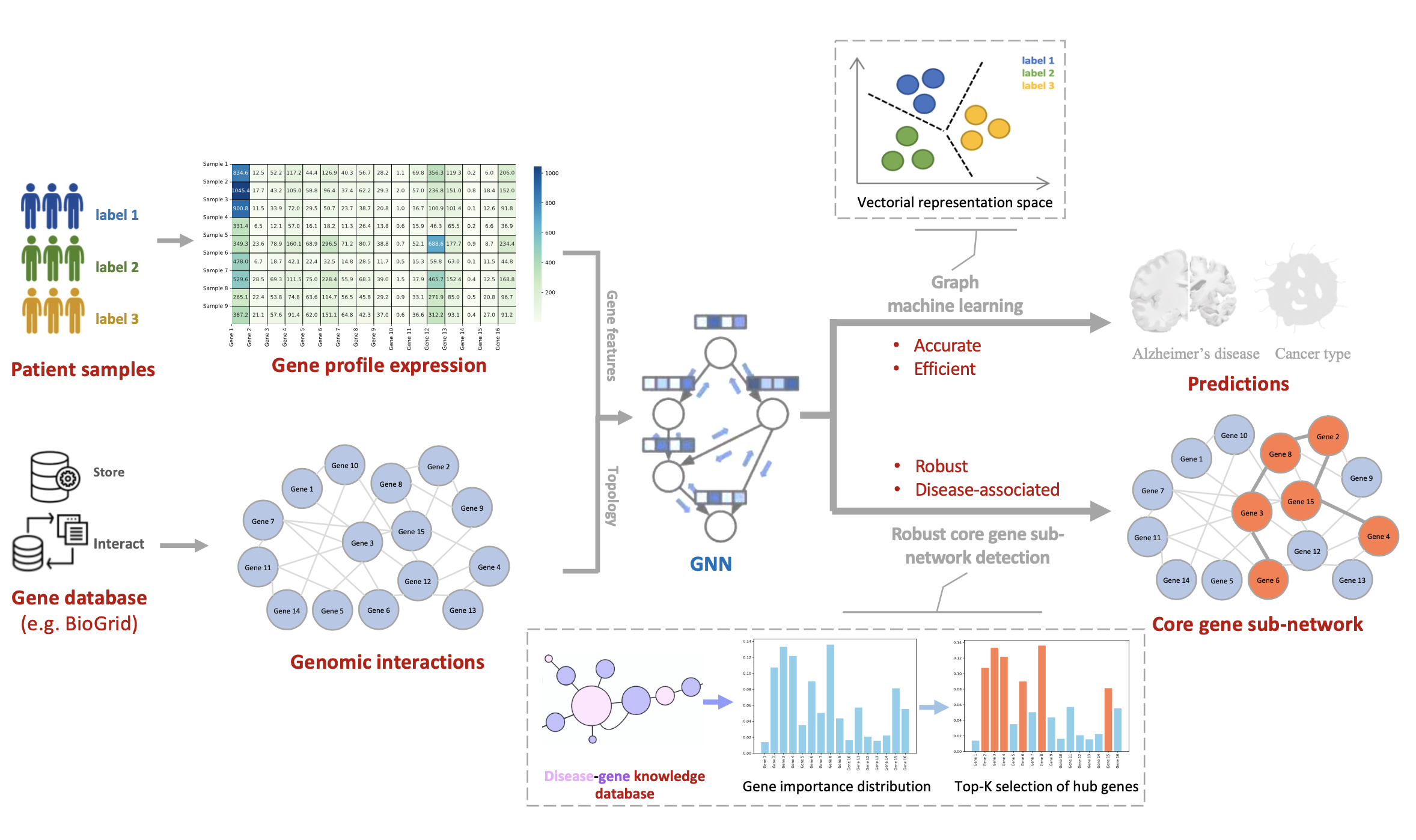}}
%\vspace{-10pt}
\caption{Overview of proposed framework for gene signaling network analysis with GNNs. Basically, gene interactions and gene expressions are obtained from genomic omics data to formulate gene networks/graphs. Then GNNs are used to perform the prediction task accurately and efficiently, while detecting robust disease-specific gene subset to understand the relation between hub genes and disease phenotypes.}
\vskip -0.18in
\label{fig:1}
\end{center}
\end{figure}

The objectives of the paper are to develop a powerful GNN for precise prediction and robust gene subset detection in gene-network-based bioinformatical tasks (see Figure ~\ref{fig:1}). Herein, we propose a novel graph convolution architecture called PathFormer encoder layer, and Figure ~\ref{fig:2}-a illustrate the architecture. The PathFormer encoder layer is constructed upon the Transformer architecture \cite{vaswani2017attention} to aggregate information through the self-attention mechanism, which is proven to be an effective solution to the over-squashing problem. Then, we use the universal orders of genes as positional encodings in the Transformer architecture, and it has been proven in \cite{Dong2023RethinkingTP} to be an ideal solution to maximize the expressive power for better prediction performance without intriguing any complexity issue. In the end, the original Transformer model is designed for sequence data rather than graphs, thus ignoring pathway information in graphs. Hence, the proposed PathFormer encoder layer injects the gene-pathway information as a learnable attention bias matrix to the attention mechanism, so that it provides flexibility to capture the pathway relation between any pair of genes in a gene network. 

Beside the expressivity to generate accurate prediction, interpretability (transparency) is also critical for deep learning models in practical bioinformatics. Thus, interpretation technique is desirable for discovering biologically meaningful pathways and processes associated with particular disease. In this paper, we aim to detect the gene subset that can be used to decipher the disease-specific biological knowledge, and develop a Knowledge-guided Disease-specific Sortpool layer (KD-Sortpool) to achieve this goal. KD-Sortpool resorts to the sort-and-pool strategy \cite{zhang2018end, lee2019self}, which sorts genes in the input gene network according to some metric value V and then keep the top K genes. To incorporate disease-specific information and biological knowledge, we use quantitative measurement, such as the GDA score from DisGeNET, to characterize the gene-disease association. Then, KD-Sortpool computes a distribution of the gene selection by normalizing the metric value V across genes, then the distribution is used to compute the expectation of the gene-disease association measurement, which is used as a regularization term in the objective function of the optimization problem. We tested our PathFormer model on two challenging bioinformatical tasks in real world: Alzheimer's disease (AD) classification task and cancer classification task. Two AD datasets (Mayo and Rosmap) and one cancer dataset are used. Experimental results demonstrate that our PathFormer model can beat existing AI models on all datasets. The average improvement of prediction accuracy is at least 38$\%$ on AD datasets and 23 $\%$ on cancer dataset.   Furthermore, we show that a disease-specific stable set of genes were identified by the PathFormer model.

\section{Methodology}
\label{back}
%\section{Conclusion and Future Work}
\subsection{Overview of the PathFormer Model}

\begin{figure}[t]
\begin{center}
%\vskip -0.15in
\centerline{\includegraphics[width=0.99\textwidth]{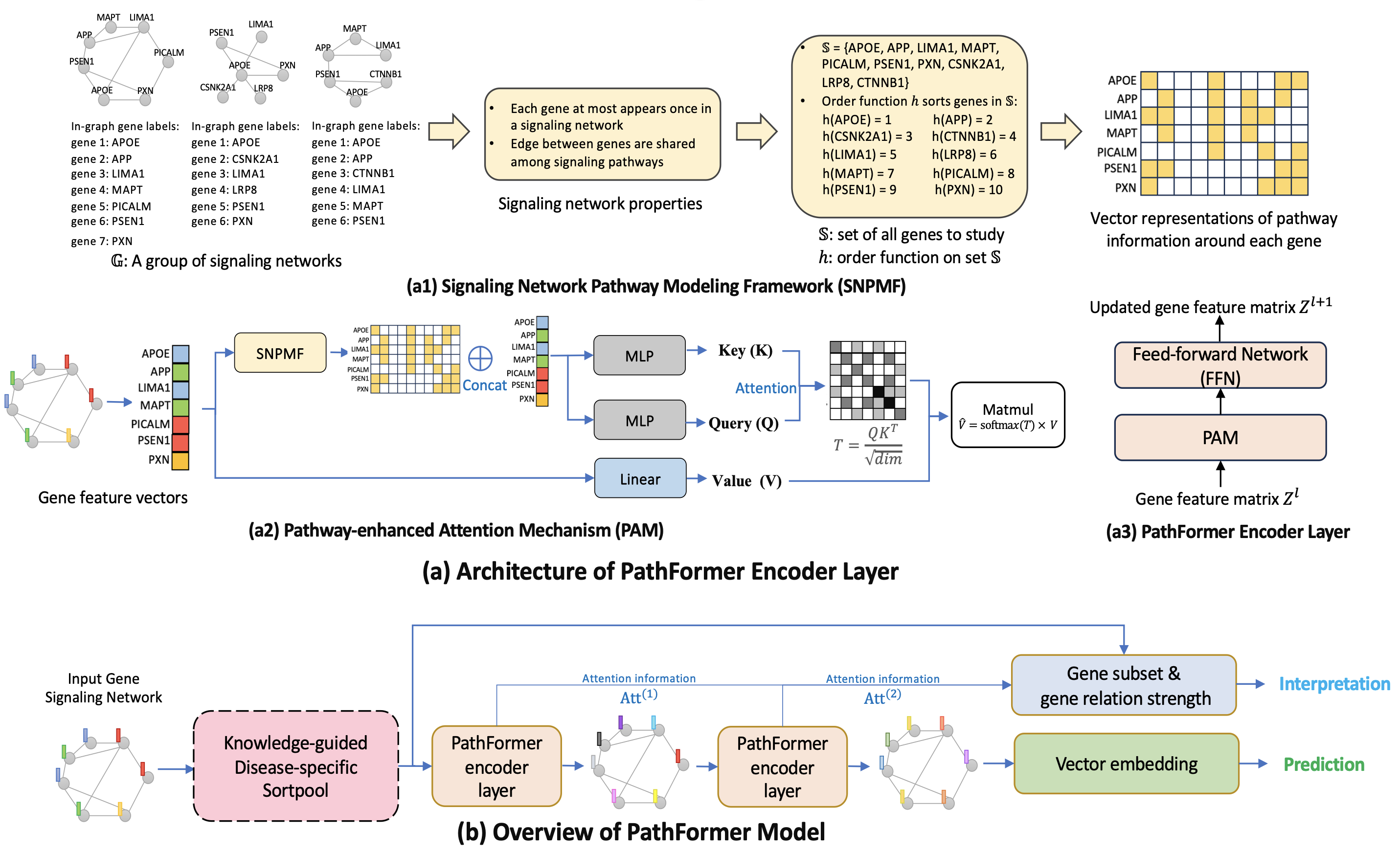}}
%\vspace{-10pt}
\caption{Architecture overview. \textbf{a.} introduces the proposed PathFormer encoder layer. PathFormer encoder layer consists of a Pathway-enhanced Attention Mechanism (PAM) and a subsequent feed-forward network (FFN). Compared to a standard attention mechanism, PAM utilizes the proposed SNPMF (Signaling Network Pathway Modeling Framework) to generate vector embeddings of pathways around each gene, which are then concatenated with gene features to compute the key matrix and query matrix. \textbf{b.} illustrates the overall architecture of the PathFormer model. PathFormer model is composed of a knowledge-guided disease-specific Sortpool (KD-Sortpool) layer and a stack of PathFormer encoder layers. It takes gene network of patients as input and outputs predictions of disease/ phenotype as well as gene subset for biological interpretations. }
\vskip -0.18in
\label{fig:2}
\end{center}
\end{figure}

Figure ~\ref{fig:2}-b illustrates the overall architecture. Specifically, PathFormer consists of a Knowledge-guided Disease-specific Sortpool layer (KD-Sortpool layer) and several PathFormer encoder layers.  Based on the prior knowledge of a particular disease, the KD-Sortpool layer select top K genes as the gene subset for the purpose of outcome interpretation. Then PathFormer encoder layers iteratively update features of each gene by aggregating its’ neighbors’ gene features. In the end, PathFormer summarize all gene features through a MLP (multiple layer perceptron) to generate a vector embedding of the input gene network, and then the vector embedding is used to predict the disease or a particular phenotype.

\subsection{Knowledge-guided Disease-specific Sortpool}

The sort-and-pool strategy is widely adopted to select “important nodes” in graphs. Existing GNN models, such as Sortpool \cite{zhang2018end} and SAGpool \cite{lee2019self}, sort nodes according to the learnt node representations/vectors after multiple graph convolutional operations. Hence, the sorting operation is individualized. That is, graphs with different node features and topology may select completed different nodes of interest. In omics data analysis studies, biomarker ranking is different, and is usually based on a group of samples like AD or cancer subtypes vs control. To achieve this purpose, we design the Knowledge-guided Disease-specific Sortpool (KD-Sortpool). 

Unlike general graphs in popular graph deep learning tasks, gene signaling networks have some unique properties that can be utilized to design powerful neural architectures. Overall, we find that each gene (name) appears at most once in each signaling network, and the connection of any pair of genes is shared among different signaling networks. Let $\mathbb{G} = \{ \mathcal{G}_{n} = (\mathcal{V}_{n},\mathcal{E}_{n}) | n = 1,2,..., N\}$ represents the group of signaling networks for all patients (AD or control samples in AD datasets) to study, where $\mathcal{V}_{n}$ contains genes of the sample $n$ and $\mathcal{E}_{n}$ contains reported connections of these genes. Then there is a gene set $\mathbb{S} = \bigcup_{\mathcal{G}_{n} \in \mathbb{G}}\{v| v \in \mathcal{V}_{n}\}$ that contains all possible genes to analyze in the group $\mathbb{G}$.

Then we can define an order function $\textit{h}$ on the overall gene set $\mathbb{S}$, such as a lexicographical sort of gene names, to provide a unique way to order genes in $\mathbb{S}$. After that, we define a learnable vector $\lambda$ of size $|\mathbb{S}|$ to formulate distribution of gene selection. In this vector $\lambda$, each element $\lambda_{p}$ assigns gene $p \in \mathbb{S}$ a trainable importance score. Consequently, for each sample/patient-specific signaling network $\mathcal{G}_{n} = (\mathcal{V}_{n},\mathcal{E}_{n}) \in \mathbb{G}$, the probability of selecting a node/gene v is computed as:

\begin{align}
\label{func:1}
    \epsilon(v) = \frac{\lambda_{h(v)}}{\sum_{u \in \mathbb{S}} \lambda_{h(u)}}
\end{align}

The function ~\ref{func:1} indicates that the gene selection process is independent of initial gene features (i.e. gene expressions). The reason is that the group-based analysis of expressions of the same gene in different sample groups (like AD sample group v.s. control sample group) sometimes provide contradictory results in different datasets. Appendix ~\ref{sec:appenF} takes the AD classification as an example, and we compare the difference of the mean gene expression between AD samples and control samples for each gene in signaling network. For most genes, the t value is significantly larger than 0 on dataset MAYO when it is significantly smaller than 0 on dataset ROSMAP, and vice versa. This observation indicates that if a disease-specific group-based gene ranking function is dependent on gene expression values, it tends to detect different patterns of important genes in dataset MAYO and ROSMAP, which leads to contradictory conclusion since both datasets are used to study the Alzheimer's disease.

To incorporate disease-specific information in the gene selection process without using gene expressions, KD-Sortpool proposes to quantify the gene-disease association of genes in the set S through prior biological knowledge. In this study, we use the open-source database DisGeNET \cite{pinero2021disgenet,kanehisa2000kegg} to simplify the quantitative estimation of each gene-disease association from text profiles of publications supporting that association. Multiple scores, such as GDA score and VDA score, are available from DisGeNET to achieve this purpose. Then, KD-Sortpool takes the GDA score of a gene v as the quantitative estimation $e(v)$ of its gene-disease association. 

Equipped with the gene selection distribution  $\epsilon(v)$ and the gene-disease association estimation $e(v)$, KD-Sortpool will deterministically sort the probabilities $\{\epsilon(v) | v \in \mathcal{V}_n\}$ of all genes in the input gene network $\mathcal{G}_{n} = (\mathcal{V}_{n},\mathcal{E}_{n})$ of a patient, then it keeps the top K genes as the gene subset of interest. Let $\mathcal{S}_n \subseteq \mathcal{V}_n$ be the gene subset. Since genes are selected in an independent way, KD-Sortpool ignores the association of multiple genes and the disease. Then, the expectation of the association between the selected gene subset $\mathcal{S}_n$ and the disease can be estimated as:
\begin{align}
    A(\mathcal{S}_n) = \sum_{v \in \mathcal{S}_n} \epsilon(v) e(v)
\end{align}

\subsection{PathFormer Encoder Layer}

Since KD-Sortpool ignores the co-effect of genes’ functionality, the following graph convolution layers (i.e. PathFormer encoder layers) are supposed to produce an output/prediction that focuses on different genes in the input signaling network. As a fundamental component in many deep neural architectures, the attention mechanism helps in identifying the most important items in the inputs and has achieved the state-of-the-art predictive performance in various deep learning tasks like natural language processing. In addition, the attention mechanism has also been proven to be an effective solution to the over-squashing problem \cite{alon2020bottleneck}. Consequently, we resort to the attention mechanism when designing the PathFormer encoder layer to improve the diagnosis accuracy.

In the context of deep learning, especially in models like Transformer, the attention mechanism assigns different weights to different items of the input by computing a key vector, query vector and value vector for each item based on its’ features. Query vectors and key vectors are used to compute the similarity scores through a similarity function like dot product, then a softmax function transforms these similarity scores into attention weights, which are used to calculate a weighted sum of corresponding value vectors to generate the outputs of the attention mechanism. Thus, when the attention mechanism is applied to graphs like signaling networks, the key and query of a node (gene) in the input graph (signaling network) is simply dependent on its’ features (profile expression). Then, the weight of a pair of nodes (genes) assigned by the attention mechanism is always the same regardless of how these two nodes (genes) are connected in the input graph (signaling network). In the field of bioinformatics and system biology, genes are studied in biological pathways to explain their relation to specific phenotype and disease. Hence, we propose a \textbf{P}athway-enhanced \textbf{A}ttention \textbf{M}echanism (\textbf{PAM}) in the PathFormer encoder layer to incorporate the pathway information of genes in the computation of their connection strength. Enumerating all pathways between each gene pair in a signaling network can be computational complex and impractical. In contrast, our PAM introduces a \textbf{S}ignaling \textbf{N}etwork \textbf{P}athway \textbf{M}odeling \textbf{F}ramework (\textbf{SNPMF}) to provide a vector for each gene in a signaling network that can injective represents upper-bounded size pathways contain the gene.

One interesting property of signaling networks is that the connection/edge of any pair of genes is shared among different signaling networks. Recall that $\mathbb{G} = \{ \mathcal{G}_{n} = (\mathcal{V}_{n},\mathcal{E}_{n}) | n = 1,2,..., N\}$ represents the group of signaling networks to study, and $\mathbb{S} = \bigcup_{\mathcal{G}_{n} \in \mathbb{G}}\{v| v \in \mathcal{V}_{n}\}$ that contains all possible genes in $\mathbb{G}$. Thus, for any two genes $u, v \in \mathbb{S}$, if they are connected in one signaling network (i.e. $\exists i$ st. $ (u,v) \in \mathcal{E}_i$), then they are also connected in any other signaling network when they are obtained in that network (i.e. for $\forall j$, we have $(u,v) \in \mathcal{E}_j$ if $u, v \in \mathcal{V}_j$). This property indicates that encoding multiple pathways is equivalent to encode genes in these pathways regardless of the gene connections. Then our objective is to find a way to encode genes that can be shared among signaling networks. The sorting function h on the overall gene set $\mathbb{S}$ provides an ideal solution as it generates the same feature for the same gene across different signaling networks. Based on above analysis, we propose SNPMF and Figure 2-a-a1 illustrates the framework. SNPMF generates a vector $p(v)$ of size $B \times |\mathbb{S}|$ to represent pathways around each gene $v$ in the signaling network $\mathcal{G}_n$. Here $B$ is a hyper-parameter that determines the bounded size of pathways. $p(v)$ is initialized initialized as a zero vector. Then, for any gene $u$ in signaling network $\mathcal{G}_n$ that is on a pathway contains gene $v$ and the distance to gene $v$ is $d < B$, SNPMF sets the element $(d-1) \times |\mathbb{S}| + h(u)$ in the vector $p(u)$ to be $1$.

Next, we introduce the Pathway-enhanced Attention Mechanism (PAM) and corresponding PathFormer encoder layer. \textbf{Figure ~\ref{fig:2}-a-a2,a3} illustrate their architectures. Compared to a standard attention mechanism, PAM concatenates vector $p(v)$ generated by SNPMF and gene features, and then utilizes MLPs to compute keys and queries of genes. Thus, the attention weights in PAM are enhanced by incorporating pathway information in the signaling network. On the other hand, a linear projection layer is used to learn the values (of genes) based on initial gene features. In analog to a standard Transformer encoder layer, our PathFormer encoder layer consists of a PAM and a subsequent feed-forward network FFN, which consists of a standard Dropout layer $\to$ FC (fully connected) Layer $\to$ Activation Layer $\to$ Dropout Layer $\to$ FC Layer, with a residual connection from before the first FC layer to after the dropout immediately following the second FC layer.

In the end, we present the overall mathematical formulation of the proposed PathFormer encoder layer. A gene subset $\mathcal{S}_n$ and $K$ genes in the input gene network will be extracted after KD-Sortpool layer. Let $Z^l = [z^l_1, z^l_2, ..., z^l_K]^T \in \mathbb{R}^{K \times d_l}$ be the matrix of gene feature vectors to the $l$-th PathFormer encoder layer, where $d_l$ is the dimension of gene features in the layer. Let $p^l = [p(v_1), p(v_2),..., p(v_K)]^T \in \mathbb{R}^{K \times B|\mathbb{S}|}$ be the matrix pathway vectors generated by SNPMF for genes in subset $\mathbb{S}$. Then, the key matrix $K^l$, query matrix $Q^l$ and value matrix $V^l$ in PAM is computed as following,
\begin{align}
    Q^l &= \textit{MLP}^l_Q (\textit{concat}(Z^l, P^l)) \\
    K^l &= \textit{MLP}^l_K (\textit{concat}(Z^l, P^l)) \\
    V^l &= \textit{MLP}^l_V (\textit{concat}(Z^l)) 
\end{align}

Then the attention matrix $\textit{Att}^l$ is computed as a softmax of the dot product of key matrix$K^l$ and query matrix $Q^l$, where the attention weights in matrix $\textit{Att}^l$ are used to calculate a weighted sum of the corresponding values in $V^l$ to capture the relevant information from the genes based on the importance indicated by the attention weights. 
\begin{align}
    \textit{Att}^l &= \frac{Q^l (K^{l})^{T}}{\sqrt{d_l}} \\
    \hat{V}^l &= \textit{softmax}(\exp{\textit{Att}^l})V^l
\end{align}

After that, the output of the current PathFormer encoder layer l is computed through the feed-forward network.
\begin{align}
    O^l = \textit{FFN}(\hat{V}^l)
\end{align}
The output gene feature matrix $O^l$ is used as the input to the next PathFormer encoder layer $l+1$. That is $Z^{l+1} = O^l$. In the first PathFormer encoder layer, the input feature vector of a gene $v$ takes the concatenation of its’ expression values and the one hot encoding of $p(v)$, where the one hot encoding of $p(v)$ works as the positional encoding in a standard Transformer model to identify genes' position based on it’s order in the set $\mathbb{S}$. 

\subsection{Readout Mechanism}

The last PathFormer encoder layer (i.e. layer $L$) can output matrix $O^L$ (or $Z^{L+1}$) that contains learnt embeddings of genes (i.e. gene feature vectors) in the gene subset $\mathcal{S}_n$ generated by KD-Sortpool. We seek a readout mechanism to generate a vector $z$ from $O^L$ as the representation of the input gene signaling network. To avoid the information loss, the readout mechanism needs to encode the order of genes in the universal gene set S and contain all genes in the gene subset $\mathcal{S}_n$. Thus, we use,
\begin{align}
    z = \sum^{K}_{k=1} W_{h(v_k)} O^L_k
\end{align}
where $W_{h(v_k)}$ is the trainable weight matrix related to gene whose order in set $\mathbb{S}$ is $h(v_k)$. In the end, $z$ is submitted to a MLP for obtaining the final prediction $\hat{y}$, which can provide the estimated probability vector of classification through a softmax operation.

\subsection{Loss Function}
\textbf{Cross Entropy (CE) Loss} This paper studies the Alzheimer’s disease (AD) classification and cancer subtype classification, thus the classification loss takes the cross-entropy loss,
\begin{align}
    \textit{L}_{ce} = \frac{1}{N} \sum^{N}_{n=1} \sum^{C}_{c=1}-y_{n,c}\log{\hat{y}_{n,c}}
\end{align}
Where $N$ is the number of samples/patients; $C$ is the number of classes in the problem; $y_(n,c)$ is the ground truth label of patient/sample $n$ such that $y_(n,c)=1$ if the sample is in the class $c$; $\hat{y}_{n,c}$ is the $c$-th entry of $softmax(\hat{y})$ of the patient/sample $n$.

\textbf{Gene Subset Consistency (GSC) loss}. KD-Sortpool introduces an approach to estimate the association strength $A(\mathcal{S}_n)$  between the gene subset $\mathcal{S}_n$ and a particular disease based on the trainable distribution $\epsilon(v)$ and the gene-disease association value $e(v)$. To force the gene selection process to be disease-specific and consistent with prior biological knowledge associated with the disease of interest, we propose the gene subset consistency (GSC) loss based on the formulation of $A(\mathcal{S}_n)$. Since $e(v)$ takes the GDA score of gene, which ranges from 0 to 1, $A(\mathcal{S}_n)$ is upper bounded by $\sum_{v \in \mathcal{S}_n} \epsilon(v)$ . Then, our GSC loss takes:
\begin{align}
    \textit{L}_{gsc} = \frac{1}{N}\sum^{N}_{n=1}\sum_{v \in \mathcal{S}_{n}}\epsilon(v) (1 - e(v))
\end{align}
Thus, a smaller GSC loss indicates a stronger association of the selected gene subset $\mathcal{S}_n$ and the disease of interest. Lastly, the overall loss function takes following formulation,
\begin{align}
    \textit{Loss} = \textit{L}_{ce} + \sigma \textit{L}_{gsc}
\end{align}

Where $\sigma$ is a tunable hyper-parameter. The objective is to minimize the objective loss. When samples from diseases of significantly different prior biological knowledge, $\textit{L}_{gsc}$ serves as a regularization term and will penalize it if the KD-Sortpool select very similar gene subsets for these completely different diseases. 

\subsection{Interpretation from PathFormer }

\textbf{Target identification and target-target co-effect estimation.} The remaining genes in the last graph layer are the identified targets. The number of remaining genes, like $50$ or $100$ targets, is a model parameter that is set by users. In analog to self-attention mechanism, the Pathway-enhanced Attention Mechanism (PAM)  in the PathFormer encoder layer provides an in-hoc approach to interpret the co-effects of genes. The attention matrices $\{\textit{Att}^{l,n} | l = 1,2,..., L; n = 1,2, ...N\}$ enable us to compute the population-based connection strength between any gene pair $(i,j)$ as following:
\begin{align}
    \alpha_{i,j} = \frac{1}{NL} \sum^{N}_{n=1}\sum^{L}_{l=1} \textit{Att}^{l,n}
\end{align}
Here $N$ is the number of patients/samples; $L$ is the number of PathFormer encoder layers.

%\section{Proposed Model}
%\input{3_method}

\section{Experiments}
\label{exper}
\subsection{Datasets and Metrices}

\textbf{Alzheimer’s disease datasets (Mayo and Rosmap):}Two datasets, Mayo and Rosmap, are used as the benchmark datasets of Alzheimer’s disease prediction problem in bioinformatics \cite{allen2016human,de2018multi}. The objective is to distinguish Alzheimer’s disease (AD) samples from normal elderly controls \cite{custodio2022functional}.  Mayo dataset is composed of control tissue samples and AD pathological aging samples, while ROSMAP dataset contains control samples and AD dorsolateral prefrontal cortex samples. 

The gene features in Mayo and RosMap are first mapped to the reference genome using STAR (v.2.7.1a), and then the transcriptomic (TPM) values of $16,132$ common protein-coding genes are obtained in both datasets by applying the Salmon quantification tool in alignment-based RNA-seq data. The Mayo dataset contains $158$ graphs, each including 3000 genes, while the Rosmap dataset contains 357 graphs, each also including $3000$ genes. The edges between genes are collected from KEGG (Kyoto Encyclopedia of Genes and Genomes) database \cite{kanehisa2000kegg} based on the physical signaling interactions from documented medical experiments. According to the Biological General Repository for Interaction Datasets (BioGRID: \url{https://thebiogrid.org/}), any two interrelated genes are undirected.

\textbf{Cancer datasets (Cancer): } To understand differences in biological mechanisms among cancer subtypes, we design the Cancer dataset. Cancer dataset aims to predict the type of caner samples based on the gene network structure and gene features. Gene features and cancer labels are extracted from the Xena server: \url{https://xenabrowser.net/.} The edges between genes are also collected from KEGG. Patient samples are collected from the longevity dataset. This dataset contains 18 different typical cancer types, including uterine carcinosarcoma, thyroid carcinoma, acute myeloid leukemia, skin cutaneous melanoma, thymoma, testicular germ cell tumor, stomach adenocarcinoma, sarcoma, rectum adenocarcinoma, prostate adenocarcinoma, pancreatic adenocarcinoma, ovarian serous cystadenocarcinoma, lung adenocarcinoma, liver hepatocellular carcinoma, mesothelioma, kidney clear cell carcinoma, head $\&$ neck squamous cell carcinoma, uterine corpus endometrioid carcinoma.

\subsection{Experiment Setup}
We use NVIDIA Tesla GTX 1080Ti GPUs to train/test our PathFormer model and other deep learning baselines. Python environment is set up and model architectures are constructed based on Pytorch and Pytorch geometric library.  To provide robust evaluation, we perform 5-fold cross validation to test the predictive performance of each model and report the average prediction accuracy as well as the standard deviation across folds. 

In the experiment, our PathFormer model is implemented with one KD-Sortpool and two subsequent PathFormer encoder layers. In the KD-Sorpool layer, we test different K (the number of gene to select) from the set $\{100, 500, 1000\}$ in the section 3.5 to validate its ability to detect gene subset of different size. In each PathFormer encoder layer, the dimension of gene features $d_l$ is set to be $32$; GNNs that computes the query matrix and key matrix takes two GIN \cite{xu2018how} graph convolution layer, where the feature dimensions of both the hidden layer and output layer are set to be $32$. Other MLPs in PathFormer take $2$ layers where the feature dimension of the hidden layer is set to be $64$. When optimizing the parameters of PathFormer model and other deep learning baselines, we use the Adam optimizer with an initial learning rate of $0.001$ and the learning rate will anneal to half every $30$ epochs; The training process is stopped when the validation metric does not improve further under a patience of $5$ epochs.

\begin{table*}[t]
\vskip -0.1in
\begin{center}
%\Large
\resizebox{0.9\textwidth}{!}{
\begin{tabular}{lcccccccc}
\toprule
& \multicolumn{2}{c}{Mayo} & & \multicolumn{2}{c}{RosMap} & &  \multicolumn{2}{c}{Cancer} \\
\cmidrule(r){2-3}  \cmidrule(r){5-6} \cmidrule(r){8-9} 
Methods  & Accuracy $\uparrow$ & F1 score $\uparrow$ & & Accuracy $\uparrow$ & F1 score $\uparrow$ & & Accuracy $\uparrow$ & F1 score $\uparrow$   \\
\midrule
\color{yellow}GIN & 0.496 $\pm$ 0.042 & 0.484 $\pm$ 0.036 & & 0.471 $\pm$ 0.039 & 0.482 $\pm$ 0.041 & & 0.537 $\pm$ 0.045 & 0.512  $\pm$ 0.047\\
\color{yellow}GCN & 0.561 $\pm$ 0.049 & 0.535 $\pm$ 0.021 & & 0.520 $\pm$ 0.036 & 0.571 $\pm$ 0.032& & 0.593 $\pm$ 0.039 & 0.561  $\pm$ 0.042\\
\color{yellow}GAT& 0.515 $\pm$ 0.034 & 0.547 $\pm$ 0.027 & & 0.491 $\pm$ 0.037 & 0.508 $\pm$ 0.042 & & 0.461 $\pm$ 0.039 & 0.532  $\pm$ 0.031\\
\color{green}Sortpool & 0.521 $\pm$ 0.034 & 0.501 $\pm$ 0.021 & & 0.522 $\pm$ 0.037 & 0.508 $\pm$ 0.042 & & 0.661 $\pm$ 0.032 & 0.642  $\pm$ 0.037\\
\color{green}SAGpool & 0.506 $\pm$ 0.047 & 0.491 $\pm$ 0.040 & & 0.491 $\pm$ 0.046 & 0.488 $\pm$ 0.036 & & 0.522 $\pm$ 0.040 & 0.507  $\pm$ 0.037\\
\color{green}Diffpool & 0.529 $\pm$ 0.031 & 0.522 $\pm$ 0.021 & & 0.517 $\pm$ 0.038 & 0.492 $\pm$ 0.026 & & 0.578 $\pm$ 0.048 & 0.601  $\pm$ 0.051\\
\color{orange}Graphformer & 0.594 $\pm$ 0.041 & 0.601 $\pm$ 0.038 & & 0.602 $\pm$ 0.050 & 0.613 $\pm$ 0.046 & & 0.739 $\pm$ 0.044 & 0.722  $\pm$ 0.058\\
\color{orange}graphTrans & 0.513 $\pm$ 0.027 & 0.526 $\pm$ 0.033 & &  0.553 $\pm$ 0.041 & 0.527 $\pm$ 0.037 & & 0.689 $\pm$ 0.041 & 0.656  $\pm$ 0.059\\
\color{blue}SANEpool & 0.517 $\pm$ 0.033 & 0.504 $\pm$ 0.031 & & 0.509 $\pm$ 0.030 & 0.481 $\pm$ 0.043  & & 0.516 $\pm$ 0.049 & 0.532  $\pm$ 0.056\\
\color{blue}MAL-GNN & 0.551 $\pm$ 0.037 & 0.579 $\pm$ 0.046 & & 0.560 $\pm$ 0.035 & 0.584 $\pm$ 0.041 & & 0.620 $\pm$ 0.029 & 0.691  $\pm$ 0.033\\
\midrule
\textbf{PathFormer} & \textbf{0.835}$\pm$ \textbf{0.036}& \textbf{0.825 }$\pm$ \textbf{0.022} & & \textbf{0.791}$\pm$ \textbf{0.025}& \textbf{0.893}$\pm$ \textbf{0.019} & & \textbf{0.834}$\pm$ \textbf{0.011}& \textbf{0.892}$\pm$ \textbf{0.032}\\
\bottomrule
\end{tabular}
}
\end{center}
\caption{Prediction results of proposed PathFormer model and deep learning baselines. Best results are highlighted. Four types of deep learning models are used as baselines and they are visualized by different colors in the table: (1) \color{yellow}popular GNNs \color{black} that achieve leading positions are mark as \color{yellow}{yellow} \color{black}; (2) dominant \color{green}graph pooling models \color{black} for subgraph extraction are marked as \color{green}{green} \color{black}; (3)  the state-of-the-art \color{orange} graph Transformers \color{black} are marked as \color{orange}{orange} \color{black}; (4) existing powerful \color{blue}deep learning models \color{black} for analyzing gene networks in other bioinformatical tasks are marked as \color{blue}blue.}
\label{tab:table1}
\vskip -0.1in
\end{table*}

\subsection{Highly accurate prediction capability }

We compare our PathFormer model with existing state-of-the-art deep learning (DL) models and popular DL models for gene expression analysis to evaluate the classification accuracy in different bioinformatical tasks. To better demonstrate the effectiveness of the proposed PathFormer model, we select 4 types of baseline deep learning (DL) models: (1) popular GNNs that achieve top positions in various leaderboards: GIN\cite{xu2018how}, GAT\cite{Velickovic2018GraphAN}, GCN \cite{kipf2016semi}; (2) dominant graph pooling models capable of extracting ‘core subgraph’ for prediction: Sortpool \cite{zhang2018end}, SAGpool \cite{lee2019self}, Diffpool \cite{ying2018hierarchical}; (3) the state-of-the-art graph Transformers: Graphormer \cite{ying2021transformers}, graphTrans \cite{wu2021representing}; (4) powerful DL models for analyzing gene networks in other bioinformatical tasks like drug synergy prediction: MLA-GNN \cite{xing2022multi}, SANEpool \cite{dong2023interpreting}. We provide the implementation details of these baselines in Appendix ~\ref{sec:appenC}. 
  
\begin{figure}[t]
\begin{center}
%\vskip -0.15in
\centerline{\includegraphics[width=0.9\textwidth]{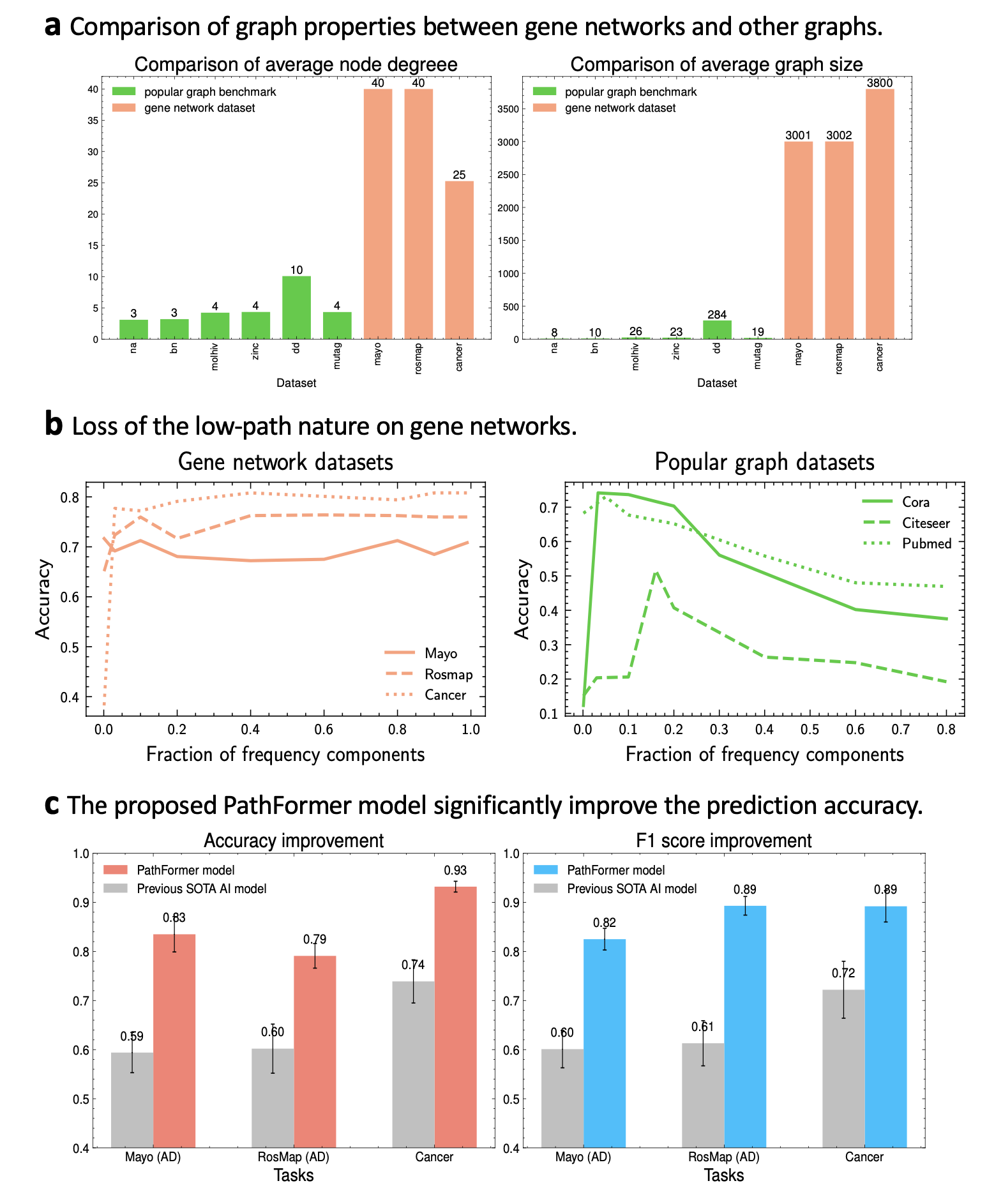}}
\vspace{-5pt}
\caption{\textbf{a} Gene networks always have significantly larger graph size and cardinality than popular graphs, which causes severe over-smoothing problem in graph machine learning.  \textbf{b} Popular graphs are always treated as signals consist of a low frequent true feature and a high frequency noise. Hence, the low-path nature indicates graph neural networks can be designed to filter out high frequency component to achieve good performance. However, gene networks do not have the low-path property.  \textbf{c} PathFormer addresses problems in a and b, thus significantly improving the prediction results over existing state-of-the-art (SOTA) deep learning models.}
\vskip -0.18in
\label{fig:Fig3}
\end{center}
\end{figure}

\textbf{Table ~\ref{tab:table1}} reports the comparison results of our PathFormer model and all baseline DL models using two evaluation metrics: classification accuracy and F1 score. Figure ~\ref{fig:Fig3}-c compares PathFormer and the best existing DL model. In the experiment, the KD-Sortpool in PathFormer model keeps all genes. The experimental results show that our PathFormer model can consistently and significantly improve the prediction result over all baselines. The average improvement of prediction accuracy is at least 38$\%$ on AD classification and 23$\%$ on cancer classification.   

Especially, previous DL models only achieve a prediction accuracy slightly better than random guess (accuracy = $0.5$) in AD classification, which limits the applicability in real world. Now, our PathFormer improves the prediction accuracy in AD classification to around $0.8$, which is a desirable level for applications.

\subsection{Highly reproducible biomarker detection}
\label{exp:predict}

\begin{figure}[t]
\begin{center}
%\vskip -0.15in
\centerline{\includegraphics[width=0.9\textwidth]{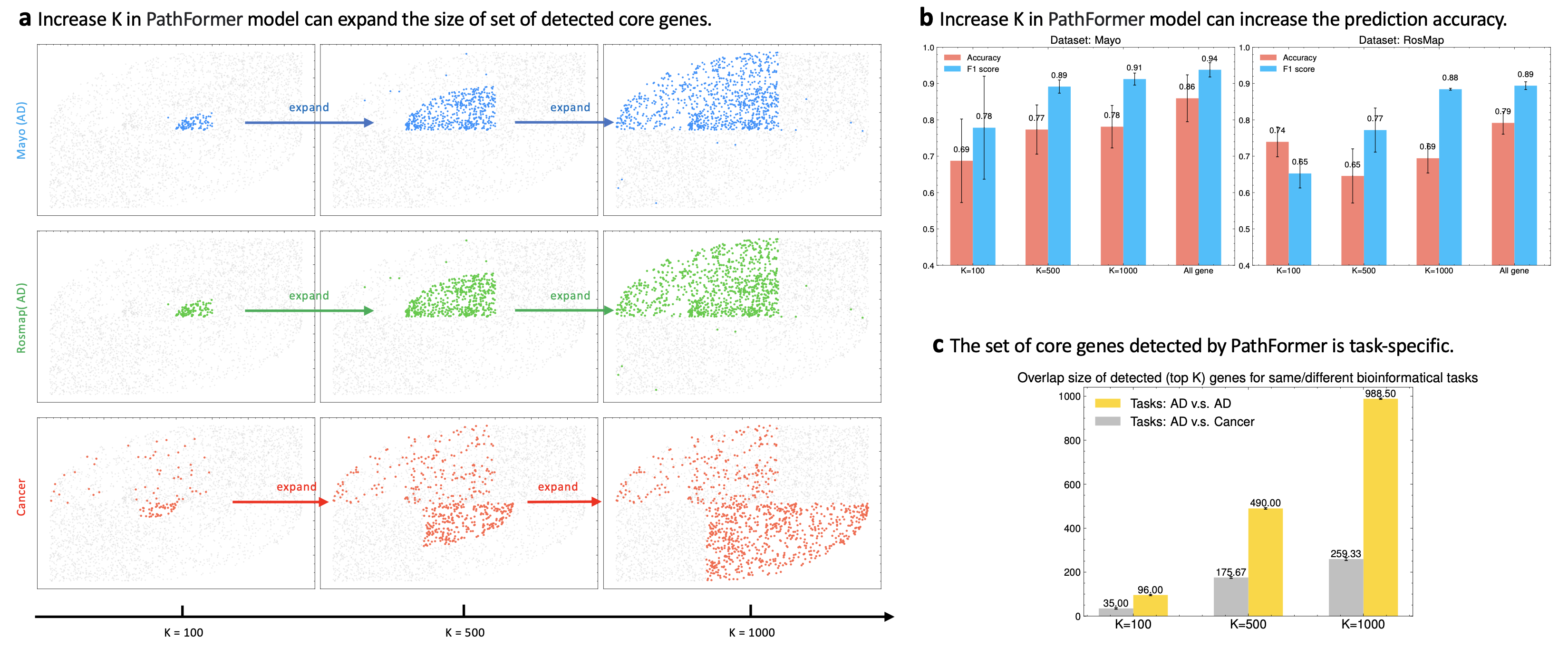}}
\vspace{-5pt}
\caption{\textbf{a} PathFormer can control the size of detected gene subset by increasing K in the KD-Sortpool layer, and the detected gene subset expands as K increases. The position of the same gene is shared among these 9 figures. \textbf{b} PathFormer can provide more accurate prediction results when K is increased to keep more genes. \textbf{c} PathFormer detects similar gene subsets for gene-network datasets of the same disease/phenotype, and different gene subsets for datasets of different diseases/phenotypes.}
\vskip -0.18in
\label{fig:Fig4}
\end{center}
\end{figure}

To evaluate the biomarker detection across different datasets, we tested three different hyper-parameter $K$ (i.e. number of genes to select) in KD-Sortpool, i.e., $K = 100$, $K = 500$, $K = 1000$, $K=$number of all genes (optional). Figure ~\ref{fig:Fig4}-a visualizes the detected gene subsets from two AD datasets (i.e. Mayo $\&$ RosMap) and one cancer dataset. We find that the detected gene subset expands as we increase K value. That is, if a gene is in the gene subset when $K = 100$, then the same gene will also appear in the detected gene subset when $K = 500$ or $K = 1000$. This property is desirable as we will not get contradictory results when using different K. If a gene is among the top $500$ important genes for a disease of interest, yet it is not among the top 100 important genes, it will confuse researchers who use the model to search gene subset of different size. Figure ~\ref{fig:Fig4}-b shows the F1 score and classification accuracy of PathFormer model on AD classification datasets. Though we observe a decrease of classification accuracy on Rosmap when K increases from $100$ to $500$, the improvement of prediction results is observed in other situations, as keeping more genes in KD-Sortpool can help reduce the information loss. Figure ~\ref{fig:Fig4}-a,c also compare the pattern of detected gene subset for different diseases. As Figure ~\ref{fig:Fig4}-a shown, no matter which K is used, the patterns of detected gene subsets from the AD datasets (i.e. Mayo $\&$ RosMap) are very similar, yet they are different from the pattern of detected gene subsets from the cancer dataset. To quantitatively describe this observation, Figure ~\ref{fig:Fig4}-c computes the overlap size of detected gene subsets for same/different disease, and we find that the overlap size is significantly large when detected gene subsets are related to the same disease/phonetype.

\section{Discussion}
\subsection{Signaling networks.}

The effectiveness of deep learning models in analyzing graphs is usually affected by graph properties. For instance, Morris et al. \cite{morris2019weisfeiler} and Xu et al. \cite{xu2018how} show that message passing GNNs cannot be more powerful than 1-dimensional Weisfeiler-Lehman (1-WL) algorithm \cite{leman1968reduction} in distinguishing non-isomorphic graphs, thus these GNNs will ignore the cyclical information and cannot provide desirable prediction results on social networks where cycles are critical features. In this section, we discuss some properties of gene networks that explain the subpar performance of existing deep learning models. 

\paragraph{Figure ~\ref{fig:Fig3}-a: extremely large node degree and graph size.} Compared to graphs in popular benchmark datasets, gene networks always have significantly larger average node degrees and graph sizes. Figure ~\ref{fig:Fig3}-a compares gene-network datasets used in this paper (Mayo, Rosmap, Cancer) and 6 popular graph datasets: NA\cite{krizhevsky2009learning}, BN \cite{krizhevsky2009learning}, molhiv \cite{dwivedi2020benchmarking}, ZINC\cite{irwin2012zinc}, D$\&$D\cite{dobson2003distinguishing}, MUTAG[41]\cite{dobson2003distinguishing}. We find that the average node degrees in graphs from popular graph datasets is usually smaller than 10, while that in gene networks are usually larger than 25. On other hand, gene networks are large-scale graphs and usually contains more than $3000$ genes, while popular graph benchmarks usually consider small-scale graphs.

Two severe consequences can be caused by above properties. (1) First, the extremely large node degrees will lead to the over-squashing problem \cite{alon2020bottleneck}, which states that the receptive field of nodes will grow exponentially with the number of GNN layers. Since the size of the receptive field reflects how much information to encode, then dominant GNNs are susceptible to a bottleneck as they aggregate too much information to a single vector, where the exponentially growing information are squeezed into fixed-size vectors. (2) Second, as popular GNNs are limited in their expressivity to encode all graph information, a large body of works is proposed to design more powerful GNNs. Basically, these works can be categorized into two groups: subgraph-based GNNs \cite{zhang2021nested} and high-order GNNs  \cite{grohe2021logic}, and the complexity of them is at least $O(n^2)$, where $n$ is the graph size. Hence, these advanced deep learning models to enhance the prediction performance will have the space/time complexity issue when applied to large-scale graphs like gene networks.

\paragraph{Figure ~\ref{fig:Fig3}-b: absence of the low-path nature.} In graph machine learning, the node features are often regarded as signals on nodes \cite{ortega2018graph}. Various prior works \cite{hoang2021revisiting,zhu2021interpreting,pan2020unified} have observed that node features of graphs in popular graph datasets consist of low-frequency true features and high-frequency noises. This property is called the low-path nature, and Figure  ~\ref{fig:Fig3}-b illustrates the property. This figure reports the average performance of $2$-layers MLPs on frequency-limited feature vectors, where node features are first transformed to the graph Fourier space and then the features vectors used for prediction is reconstructed based on top frequency components. Green curves in the right of Figure  ~\ref{fig:Fig3}-b implies that true features (i.e. low-frequency components of node features in graphs) have sufficient information for graph machine learning on popular graph datasets.

It is proven that the power of popular GNNs comes from the ability to filter out the high-frequency components in node features, and popular GNNs essentially act as a low-pass filter on graphs. However, when we perform the same experiment on gene network datasets, orange curves in the left of Figure~\ref{fig:Fig3}-b indicates that gene networks do not have the low-path nature, thus both the low-frequency components and high-frequency components are important. Thus, popular GNNs that filter out high-frequency components of node features in graphs are not suitable for gene networks.

\subsection{Theoretical Results}

In this section, we provide a mathematical formulation of graph machine learning problem, which illuminates the theoretical solution to design a GNN architecture that can address the over-squashing problem and the absence of the low-path nature. 

\paragraph{Notations} Let $\mathbb{G} = \{ \mathcal{G}_{n} = (\mathcal{V}_{n},\mathcal{E}_{n}) | n = 1,2,..., N\}$ be a set of graphs, where $\mathcal{V}_{n}$ and $\mathcal{E}_{n}$ contain nodes and edges information in graph $\mathcal{G}_{n}$. For a graph $\mathcal{G}_{n}$, each node/gene $i \in \mathcal{V}_{n}$ as a d-dimensional feature $x_i$, while the graph $\mathcal{G}_{n}$ has a label $y_n$ to predict. We use $A$ to denote the adjacency matrix, and $D$ to denote the diagonal matrix such that $D_{ii} = \sum_{j} A_{i,j}$. Then we set $\tilde{D} = D + I$.

\begin{defi}
\label{def:1}
(Optimization formulation) Let $\hat{X}$ be the output of a GNN model/layer $f$ such that $\hat{X} = f(X,A)$, where $X \in \mathbb{R}^{n \times d}$  is the input feature matrix. Let $\mathcal{N}(i) = {j \in \mathcal{V}_n | (i,j) \in \mathcal{E}_{n}}$ be the set of neighbors of node/gene $i$. Then the unconstrained optimization problem is formulated as following,
\begin{align}
    \textit{min}_{\hat{X}} \sum_{i\in V} ||\hat{X}_{i} -X_{i}||_{\tilde{D}} + \sum_{i}\sum_{j \in \mathcal{N}(i)}||\hat{X}_{i} - \hat{X}_{j}||_{2}^{2}
\end{align}
\end{defi}

Where operation $\tilde{D}$-inner product is defined as $||x||_{\tilde{D}} = (x,x)_{\tilde{D}}^{\frac{1}{2}}$. The first term in the objective of the optimization formulation constrains that output of GNN should not be too far off the input, while the second term indicates that the formulation essentially is a type of Laplacian smoothing over the whole graph, where  $p_{i,j}$ characterizes the similarity between a node/gene pair $(i,j)$.

\begin{theorem}
\label{theo:1}
The optimal solution $\hat{X}^{*}$ to the optimization formulation solves the challenge of absence of low-path nature. Let M denote the mask matrix such that $M_{i,j}=1$ if $j \in \mathcal{N}(i)$ and $M_{i,j}=0$ otherwise.  Then $MPX$ is the first-order approximation of $\hat{X}^{*}$.
\end{theorem}

We prove Theorem \ref{theo:1} in Appendix ~\ref{sec:appenD}. Basically, this theorem provides insights to design a GNN that generate first-order approximation to the optimal solution, and the key problem is how to generate the trainable parameter matrix $P$. The straightforward solution is to use the global (self-)attention mechanism such that $P_{i,j} = \frac{g(X_i, X_j)}{\sum_j g(X_i, X_j)}$, as it has been shown that the attention mechanism can help to solve the over-squashing problem at the same time [21].  Thus, theorem 1 can be used to design GNN layer that addresses the over-squashing problem and the challenge of lacking low-path nature. In the proposed PathFormer encoder layer, MP is estimated through function (3), (4), (5), (6), (7), thus it brings a significant improvement of prediction results.

\section{Conclusion}
\label{convlu}
%\section{Conclusion and Future Work}
 In this paper, we introduce an interpretable graph Transformer, PathFormer. PathFormer significantly outperforms strong baselines, including dominant GNNs, recent graph Transformers, and popular gene-netwrok-specific deep learninig models. In conclusion, PathFormer can achieve highly accurate disease diagnosis accuracy and can identify a stable set of biomarkers in the omics data-driven studies.

\newpage

\bibliographystyle{plainnat}
\bibliography{reference}

\clearpage
\appendix
\section{Transformer and Graph Transformer}
\label{sec:appenA}
\subsection{Transformer on Graphs}
\textbf{Transformer} Transformer \cite{vaswani2017attention} solves the language modeling problem \citep{dai2019transformer,al2019character,devlin2019bert,lewis2020bart} using self-attention mechanism, and improves the performance over RNN-based or convolution-based deep learning models in both accuracy and efficiency. The Transformer encoder consists of a stack of Transformer encoder layers, where each layer is composed of two sub-networks: a (multi-head) self-attention network and a feed-forward network (FFN).

let $H = (h^{T}_{1}, h^{T}_{2}, ... h^{T}_{n})$ be the input to a Transformer encoder layer. In the self-attention network, the attention mechanism takes $H$ as input and implements different linear projections to get the query matrix $K$, key matrix $K$ and value matrix $V$, Then the attention matrix $A$ is computed as following to measure the similarities, which is then used to update the representation in parallel . 
\begin{small}
\begin{align}
    A = \frac{QK^{T}}{\sqrt{d_{k}}}, \ \ \ Z = \textit{softmax}(A)V
\end{align}
\end{small}
After the self-attention network, the \textit{feed-forward network} consists of two linear transformations with a Rectified Linear Unit (ReLU) activation in between to generate the output. i.e. $O = \textit{FFN}(Z)$. The FFN is composed of a standard Dropout Layer $\to$ Layer Norm $\to$ FC (fully connected) Layer
$\to$ Activation Layer $\to$ Dropout Layer $\to$ FC Layer $\to$ LayerNorm sequence, with residual connections
from $Z$ to after the first dropout, and from before the first FC layer to after the
dropout immediately following the second FC layer.

\noindent\textbf{Transformer on graphs}
Recently, the Transformer architecture is applied to graph learning tasks to avoid structural inductive bias caused by GNNs \cite{dwivedi2020generalization, kreuzer2021rethinking, mialon2021graphit}. To incorporate the topology information of graphs, various positional encoding schemes are proposed to encodes structural relations \cite{dong2022pace} or positional information \cite{ying2021transformers} about nodes. On the other hand, other works \cite{wu2021representing,chen2022structure} are designed to embed structural similarities between nodes by learning input node features with GNNs. These graph Transformers have achieved great success and sit atop leaderboards such as the Open Graph Benchmark \cite{hu2020open}.

\section{Other Backgrounds}
\label{sec:appenB}

\paragraph{Interpretable GNN:} The interpretable GNN aims to show a transparent and understandable prediction process to humans. In other words, which parts of the input have a significant impact on the prediction? For a gene network, this could be genes, relationships between correlated genes, or a combination of both, i.e., motifs \cite{xing2022multi}. Most interpretable GNNs, such as GNNExplainer, take a local interpretable mechanism to explain the key subgraphs of each graph. One of the assumptions behind this type of interpretation is that there are input components that contribute significantly to the prediction, while the insignificant components have less impact. Such an assumption can lead to the fact that these interpretable GNNs do a poor job of discovering the important subgraphs when the genetic features or the relationships among these genes cannot be clearly distinguished. Furthermore, local interpretability treats GNNs as black boxes \cite{kovalerchuk2021survey} thus limiting human trust in the given interpretation.

\paragraph{Low-path Nature of GNN:} In general graph learning problems like semi-supervised node classification, node features $x(i)$ are often regarded as signals on nodes, and techniques in graph signal processing \cite{ortega2018graph} are then leveraged to understand the signal characteristics. Various prior works \cite{hoang2021revisiting,zhu2021interpreting,pan2020unified} assume or observe that node features $x(i)$ consist of low-frequency true features and noises. Based on the assumption, numerous GNNs are designed to decrease the high-frequency components in node features, thus essentially acting as a low-pass filter on graph signals. However, the assumption is not verified on gene networks. Figure 1 shows that gene networks do not benefit from omitting high-frequency components in signals. This means that the low-pass nature might not exist in the studied problem, and only keeping low-frequency signals might degrade the performance of GNNs due to the information loss.

\section{Details of deep learning baselines}
\label{sec:appenC}

In GNN baselines, all graph convolution layers have a feature dimension of $128$; The number of graph convolution layer is selected from the set $\{2, 3, 4\}$; The graph-level readout function is selected from the function set $\{ \textit{mean}, \textit{sum}, \textit{average}\}$. In graph pooling models, SAGpool and Diffpool take two pooling layers, where the first layer keeps 500 genes and second layer keeps $100$ genes. SortPool contains one pooling layer, which sorts genes according to the learnt embeddings and then keeps the top 100 gene embeddings as the input to a CNN model. Other parameter settings in graph pooling models follow their original papers.  In graph TransFormers, the number of encoder layers is $3$, the dimension $d_k$ is set to be $16$, the number of heads is set to be $4$. Due to the property of bioinformatical dataset, Graphormer does not perform the pre-training. In graphTrans, we use GIN with $2$ layers to extract the node embeddings.

\section{Details of Definition ~\ref{def:1} and Proof of Theorem ~\ref{theo:1}.}
\label{sec:appenD}

Given a graph $\mathcal{G} = (\mathcal{V}, \mathcal{E})$ with $|\mathcal{V}| = n$. Let $J \in \mathbb{R}^{n \times n}$ be an all-ones matrix where every element is equal to one. $p \in \mathbb{R}^{n \times n}$ is the parameter matrix whose element $(i,j)$ is $p_{i,j}$. Then, we use $p[A]$, $p[J-A$] $\in \mathbb{R}^{n \times n}$ to denote the matrices such that $p[A]_{i,j} = p_{i,j}$ if $A_{i,j} = 1$ ($p[J-A]_{i,j} = p_{i,j}$ if $(J-A)_{i,j} = 1$) and $p[A]_{i,j} = 0$ ($p[J-A]_{i,j} = 0$) otherwise. In addition, we use operation $\circ$ to denote the Hadamard product of two matrices. Then the objective function $f$ in problem \ref{equ:3} can be formulated as:
\begin{align*}
    f &= \textit{Tr} ((\hat{X} - X)(\hat{X} - X)^{T} \tilde{D}) + \hat{X}^{T}((S - M \circ p[A]))\hat{X} + \hat{X}^{T}((L - M \circ  p[J-A]))\hat{X}
\end{align*}

where $S$ and $L$ are diagonal matrix such that $S_{i,i} = \sum_{j} (M \circ P[A])_{i,j}$, $L_{i,i} = \sum_{j} (M \circ P[J-A])_{i,j}$.  
\begin{align*}
    \frac{\partial f}{\partial \hat{X}} & = \tilde{D}(\hat{X} - X) + (S + L - M \circ (p[A] + p[J-A]))\hat{X}
\end{align*}
Then, given $(\tilde{D} - (M \circ p - I))$ is invertable, the optimal solution is $\hat{X}^{*} = (I - \tilde{D}^{-1}(M \circ p - I))^{-1}X$. Then the first order approximation (through the Neuman series of a matrix) is $\tilde{D}^{-1} (M \circ p) X$   and is $\tilde{D}^{-1} (M \circ p) X$ with a constant term when when the constraints are not normalized as problem \ref{equ:3}. Since $\tilde{D}^{-1}$ can be viewed as a normalization term, the solution can be simplified as $M \circ P X$ where $P \in \mathbb{R}^{n \times n}$ is a trainable parameter matrix.

\section{Additional Interpretation Results}
\label{sec:appenE}

\begin{figure}[t]
\begin{center}
%\vskip -0.15in
\centerline{\includegraphics[width=0.99\textwidth]{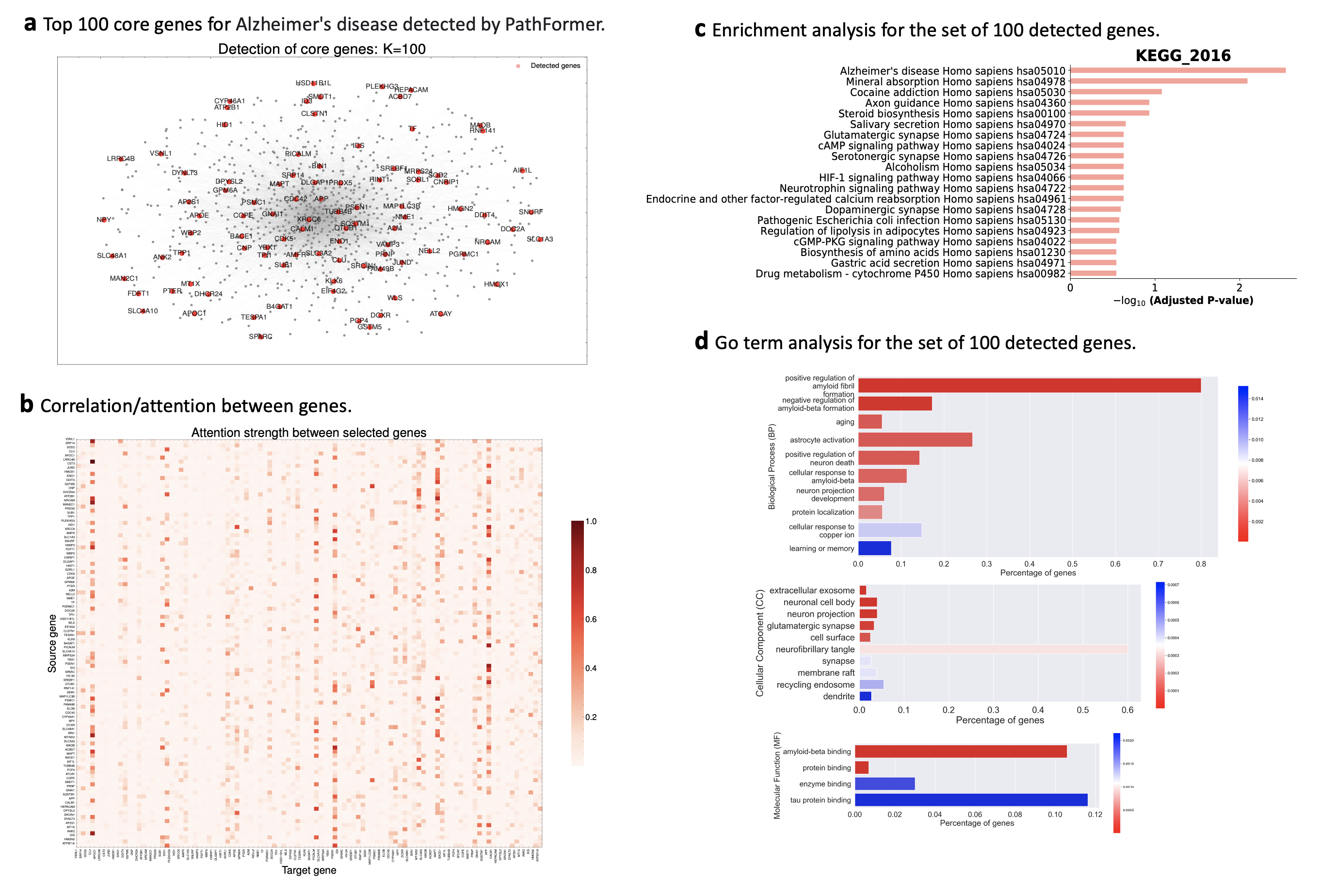}}
\vspace{-5pt}
\caption{\textbf{a} The KD-Sortpool layer in PathFormer select 100 core genes as a gene subset to explain Alzheimer’s disease.  \textbf{b} PathFormer can compute the relation strength (attention) between selected genes. Some genes (e.g. TPP1, PSEN1, CLU, APP) gain significant larger attention from other genes, and these gene usually have a large GDA score and are closed associated with Alzheimer’s disease. \textbf{c} Enrichment analysis on the detected core gene subset finds significant pathways associated with Alzheimer’s disease. \textbf{d} Go term analysis on the detected core gene subset finds significant biological process associated with Alzheimer’s disease.}
\vskip -0.18in
\label{fig:Fig5}
\end{center}
\end{figure}

The interpretability of deep learning models is of vital importance for real-world applications in the field of bioinformatics, as the interpretability can illustrate the chain of reasoning to aid in trust and to increase the testability. In aforementioned gene-network-based bioinformatical tasks, the prevailing interpretation methods aim to extract sub-networks consists of core gene pathways \cite{preuer2018deepsynergy} that reveals the underlying functional mechanism. 

\subsection{Gene set enrichment analysis. }
Gene set enrichment analysis is ubiquitous in genetic research to reveal the association of a set of genes and disease/phenotypes of interest. We implement conducted GO $\&$ KEGG to perform the gene set enrichment analysis on the detected gene subset for AD, and results are reported in c and d in Figure~\ref{fig:Fig5}.

Figure ~\ref{fig:Fig5} c illustrates the enrichment analysis results from KEGG (Kyoto Encyclopedia of Genes and Genomes) Pathways, and top 20 significant pathways (that with the smallest adjusted p-value) are presented. We find that the detected biological pathway of the greatest significance is (Alzheimer’s disease) homo sapiens hsa 05010, which belongs to the class of neurodegenerative-disease pathways and is associated with Alzheimer’s disease. 

Figure ~\ref{fig:Fig5} d presents the enrichment analysis results from GO (gene ontolog)\cite{thomas2022panther}. The p-value that characterize the significance of biological process/cellular component/molecular function is visualized as color, and we only report GO terms of a p-value smaller than $0.05$.
\begin{itemize}
    \item BP (Biological Process) results: Many reported biological processes are associated with AD, including negative regulation of amyloid-beta formation, aging, astrocyte activation, positive regulation of neuron death, cellular response to amyloid-beta, cellular response to copper ion, learning or memory. More specifically, amyloid-beta (A$\beta$) pathology constitutes an essential mechanism for AD in a person; Older age is one of the most important risk factor for AD; Astrocytes become activated around the senile plaques in AD brain; Neuronal cell death like autophagy-dependent cell death is known to be related to AD; AD can cause a faulty blood-brain barrier that prevents the clearing away of toxic beta-amyloid; Deregulated copper ions may ultimately contribute to synaptic failure, neuronal death, and cognitive decline observed in AD patients; Incapability to create new memories is often among the very first signs of AD.
    \item CC (Cellular Component) results: Many reported cellular components have been proven to be associated with AD, including extracellular exosome, neurofibrillary tangle, synapse, membrane raft, recycling endosome, dendrite. Among them, neurofibrillary tangle is related to about 0.6 percentage of genes in the detected gene subset, while others only cover a very small proportion of genes in the subset (i.e.significantly smaller than $0.1$). To explain a bit, neurofibrillary tangle is abnormal accumulations of the tau protein, while the faulty blood-brain barrier caused by AD will prevent the clearing of tau protein.
    \item MF (Molecular Function) results: In the reported molecular functions, both amyloid-beta binding and tau-protein binding are associated with AD and cover more than 0.1 percentage genes in the detected gene set. We have explained the association of tau-protein and AD in CC results, and explained the association of amyloid-beta and AD in BP results.  
\end{itemize}

\subsection{evealing relation/attention strength between genes. }
 
The attention mechanism in PathFormer encoder layers provide a built-in approach to interpret the relation/attention strength between genes in the detected gene subset using function (14). Figure \ref{fig:Fig5}-b visualizes the normalized attention strength matrix of every gene pair. We find that some target genes receive much more attentions from source genes. In the heatmap figure, we can identify these genes by searching along the x-axis for target genes whose attention vector (a line in the heatmap) contains many large values. CLU, TPP1, PSEN1, PICALM, APP are some examples. These genes usually have a relatively large GDA score, which quantities the association of a gene and AD. Here, GDA(APP) $= 0.9$, GDA(PSEN1) $= 0.7$, GDA(CLU) $= 0.5$, GDA(TPP1) $= 0.3$. Hence, these genes shows a tendency for interaction with other genes in the prediction process.

\section{Comparison of gene expressions of AD and control in different datasets }
\label{sec:appenF}

This experiment performs the independent t samples test to compare the gene expressions of AD samples and control samples in two datasets: MAYO and ROSMAP.  The objective is to show the reason why our proposed Knowledge-guided Disease-specific Sortpool (KD-Sortpool) is independent of gene expressions. In the context of bioinformatics, disease-specific biomarker/gene ranking is based on a group of samples (AD samples vs control samples). Then the ranking function will tend to select genes whose expression values are proposed to be significantly lower/higher in the AD group than the control group. Consequently, we evaluate whether the difference of mean gene expression values in AD group and control group is consistent across datasets of the same disease.

\begin{figure}[t]
\begin{center}
%\vskip -0.15in
\centerline{\includegraphics[width=0.8\textwidth]{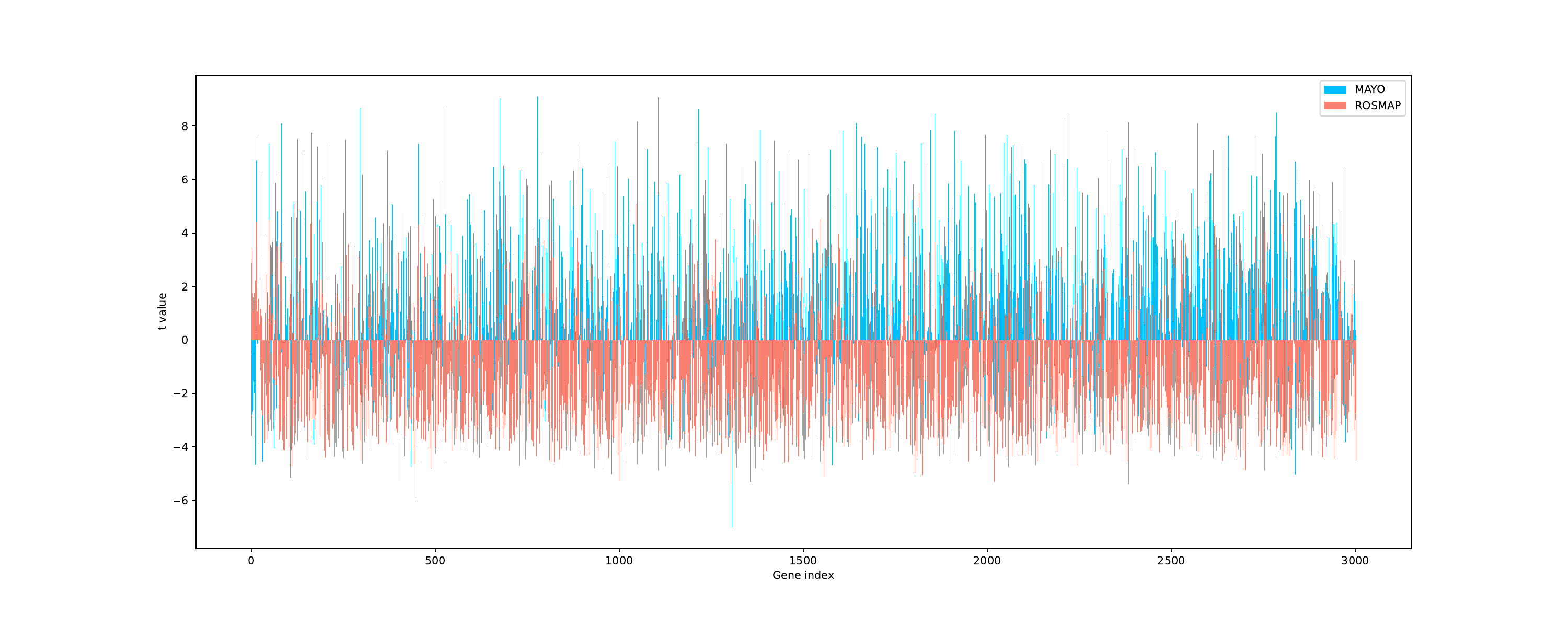}}
\vspace{-5pt}
\caption{Comparison of t statistics of each gene in dataset MAYO and ROSMAP.}
\vskip -0.18in
\label{fig:Fig6}
\end{center}
\end{figure}

Figure ~\ref{fig:Fig6} demonstrates the experimental results. We find that the t-values show different pattern in dataset MAYO and ROSMAP. For many genes, the t value is significantly larger than 0 on dataset MAYO when it is significantly smaller than 0 on dataset ROSMAP, and vice versa. This observation indicates that if a disease-specific group-based ranking function is used to select the genes of importance, it will detect different gene patterns in dataset MAYO and ROSMAP, which can lead to contradictory conclusion since both MAYO and ROSMAP are used to study the Alzheimer's disease.

\end{document}